\documentclass{webofc}

\usepackage[varg]{txfonts}   
\usepackage{hyperref}
\usepackage{url}
\usepackage{graphicx}
\usepackage{subcaption}
\hypersetup{colorlinks=true,citecolor=blue,urlcolor=blue,linkcolor=blue}
\begin{document}
\begin{flushright}
	DESY-24-161
\end{flushright}

\title{Searches for BSM physics at a gamma-gamma collider with Energy < $12$ GeV based on European XFEL}

%
%

\author{\firstname{Marten}
\lastname{Berger}\inst{1}\fnsep\thanks{\email{marten.berger@desy.de}}
\and\firstname{Gudrid}
\lastname{Moortgat-Pick}\inst{1,2}\and 
\firstname{Monika}
\lastname{Wüst}\inst{1}
}

\institute{II. Institut für Theoretische Physik, Universität Hamburg, Luruper Chaussee 149, 22761 Hamburg, Germany
\and
           Deutsches Elektron-Synchrotron DESY, Notkestr. 85, 22607 Hamburg, Germany 
          }
\abstract{The possibility of a Photon-Photon collider extension to the Beam dump of the $17.5$ GeV European XFEL has been discussed before as the first high energy collider of its sort. It would not just be to study the concept of photon colliders but would also be a collider without competition in the region of $5 - 12$ GeV for photon-photon collision. In this range, $b\Bar{b}$ and $c\Bar{c}$ resonances, tetraquarks as well as mesonic molecules can be observed. Furthermore, some BSM processes can also be reached in this range. In this paper we want to discuss the possibility of observing ALPs at such a collider. We will use a simplified description of the compton backscattering process to get a first look at cross sections and extend this to the full beam dynamics included prediction.}
%
\begin{center}
	\maketitle
\end{center}
%
\section{Introduction}
Future $e^+e^-$ linear colliders are an intensively discussed next step to address some of the most pressing questions in particle physics. They offer several unique advantages and have the potential to complement current collider projects in exploring new frontiers. One of the primary motivations for linear colliders is the study of the Higgs boson, following its discovery at the LHC in 2012 \cite{ATLAS_2012,CMS_2012}. Linear colliders, like the International Linear Collider (ILC) \cite{ILC-2019}, HALHF \cite{HALHF_2023} or the Compact Linear Collider (CLIC) \cite{CLIC_2016}, are designed to operate at energies between 250 GeV and 3 TeV. This makes them ideal for producing and analyzing the properties of the Higgs boson, like the trilinear Higgs coupling, needing a collider with at least $\sqrt{s}=500$ GeV and could lead to better understanding of the form of the Higgs potential as well as substantial further insights on the beginning of our universe.\\
The main drawback for linear colliders is that the beam can only be used once; however, this allows the opportunity to obtain high energy photons by Compton backscattering of laser light from the electron beam, giving access to $\gamma\gamma$ and $\gamma e$ interactions with luminosities and energies comparable to those of the $e^+e^-$ mode \cite{Ginzburg1,GINZBURG2,Ginzburg3,Telnov_1989}. This would extend the reach and complement the searches of the $e^+e^-$-collider, giving unique access to so far impossible or very rarely observed physics of the SM and BSM.\\
The European-XFEL is a free-electron laser that has been in operation since 2017, generating high energy X-ray beams by using a $17.5$ GeV electron linac. After passing through the undulators the electron beam is sent to the beam dump. At this point a gamma-gamma collider could be implemented, by splitting the electron beam, sending it along a curve and producing high energy photons via Compton backscattering. This idea has been proposed by V. I. Telnov \cite{Telnov_2020,TELNOV_talk}, in order to already work with the technology and difficulties of a gamma-gamma collider, developing the methods and expertise for at this time to be able to use them instantly at future colliders.\\
In the proposed range of up to $12$ GeV, the collider could already complement SM searches and even help to a better understanding of some of the most fundamental processes, like light-by-light scattering. The first observation of light-by-light scattering has been achieved by ATLAS in 2017, by colliding heavy lead ions \cite{ATLAS_LbyL}. A real photon collider would give unprecedented direct access to this process, offering a deeper understanding and chances to look for deviations from the SM.\\
Axion-like particles (ALPs) could contribute to such deviations, they are hypothetical, weakly interacting BSM particles. Initially inspired by the axion, which was proposed to solve the strong CP problem in quantum chromodynamics (QCD) \cite{PQ_1977}, ALPs generalize the idea by decoupling the mass and coupling of these particles from the QCD context. They are typically described as pseudo-Nambu-Goldstone bosons arising from spontaneous symmetry breaking at high energy scales. Heavy ALPs, MeV-TeV range, could be the mediator to yet undiscovered dark matter particles or the dark sector \cite{AxionPortal}. There are many collider searches ongoing in this mass range \cite{ColliderALPs}. Due to the ALPs coupling to the SM photon, it would also be possible to search for it at gamma-gamma colliders, by looking for deviations in the observation of light-by-light scattering.\\
Section \ref{sec:gammagamma} will discuss the concepts and most important formulas for a gamma-gamma collider as well as take a look on the energy spectrum and luminosity for unpolarized and polarized electron beams. In section \ref{sec:lbyl}, the SM light-by-light scattering process at the gamma-gamma collider is discussed, afterwards in section \ref{sec:alps} the same process will be studied with an additional ALP as an example for a further BSM collider search strategy. At the end a short conclusion and outlook is given.

\section{Gamma-gamma collider}
\label{sec:gammagamma}
\begin{figure}[htbp]
	\centering
	\includegraphics[scale=0.2]{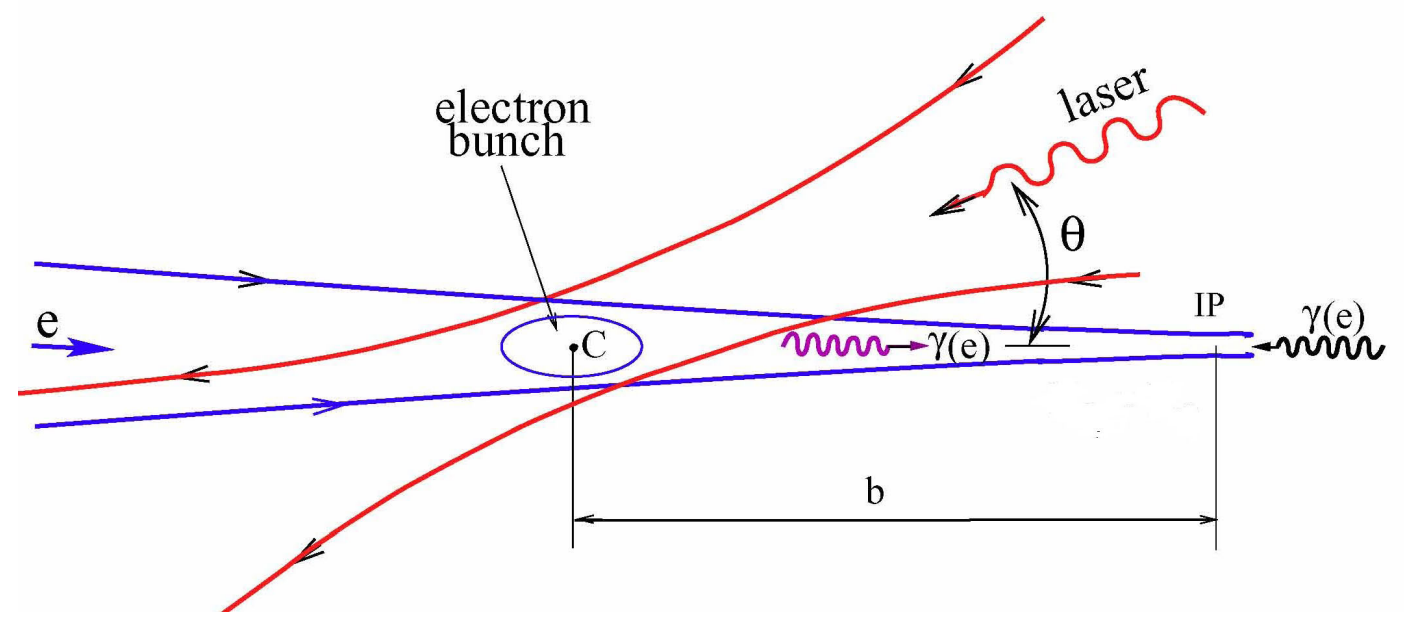}
	\caption{General scheme of a $\gamma\gamma$, $\gamma e$ collider \cite{Telnov_2020}}
	\label{fig:scheme}
\end{figure}

In order to convert a large number of electrons, the $e$-beam is collided with a laser at a conversion point (C) a small distance $b$ before the interaction point (IP) as shown in Fig. \ref{fig:scheme}. The conversion takes place via Compton backscattering, resulting in real photons. This extends the possible channels of the $e^+e^-$-collider to also include $\gamma\gamma$ and $\gamma e$ collisions, with comparable luminosities to the $e^+e^-$ collision.

At the C, a laser photon of energy $\omega_0$ collides with an electron of energy $E_0$ resulting in a high-energy photon with up to maximum energy:
\begin{equation}
	\omega_{max} = \frac{x}{x+1+\xi^2}E_0,
\end{equation}
where
\begin{equation}
	x = \frac{4E_0\omega_0}{m_e^2c^4}\cos^2\frac{\theta}{2}\simeq  15.3 \left[\frac{E_0}{\text{TeV}}\right]\left[\frac{\omega_0}{\text{eV}}\right]=19\left[\frac{E_0}{\text{TeV}}\right]\left[\frac{\mu\text{m}}{\lambda}\right],
\end{equation}
with $m_e$ the electron mass and $\xi$ the parameter describing the nonlinear effects \cite{Ginzburg_1984}. The optimal choice of parameters is $x=4.8$, leading to $\omega_{max}=0.83\text{ } E_0$ giving photons with energies close to the electron beam energy. For higher values of $x$, the Breit-Wheeler process causes a significant reduction in overall luminosity. Due to this, going beyond $x=4.8$ is unfavourable for optical laser. However, recent papers have proposed the idea of using XFELs instead of optical lasers, reaching $x$-values of up to $1000$ while still achieving high conversion rates \cite{Barklow}.

\begin{figure}[htbp]
	\begin{subfigure}[b]{0.5\linewidth}
		\centering
		\includegraphics[width=1\linewidth]{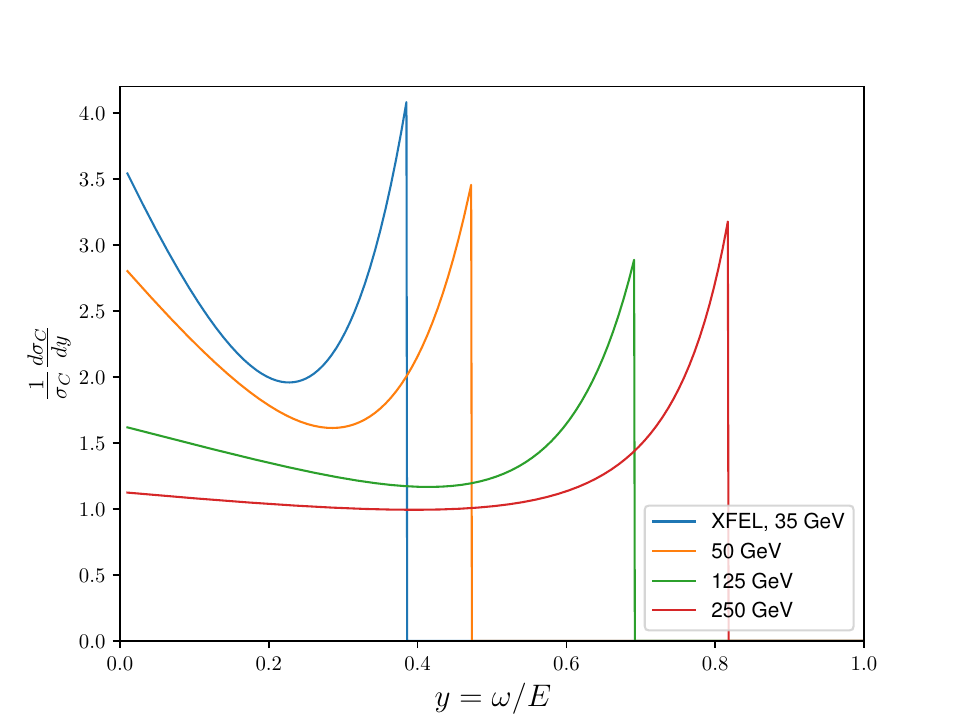} 
		\label{fig:unpolEnergy} 
	\end{subfigure}
	\begin{subfigure}[b]{0.5\linewidth}
		\centering
		\includegraphics[width=1\linewidth]{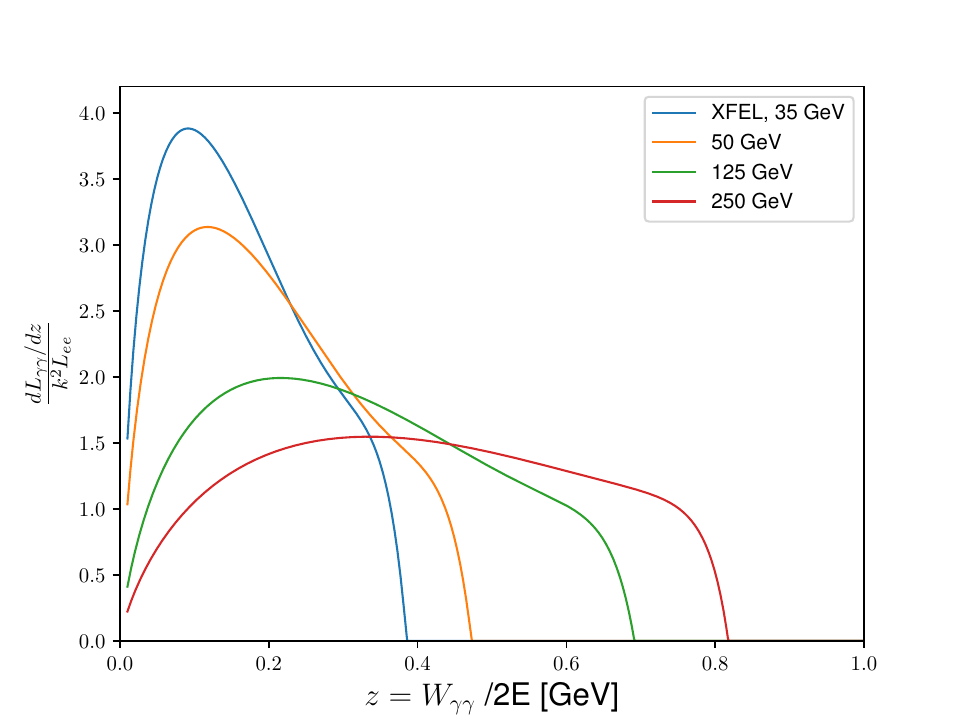} 
		\label{fig:unpolSpectrum}
	\end{subfigure} 
	\begin{subfigure}[b]{0.5\linewidth}
		\centering
		\includegraphics[width=1\linewidth]{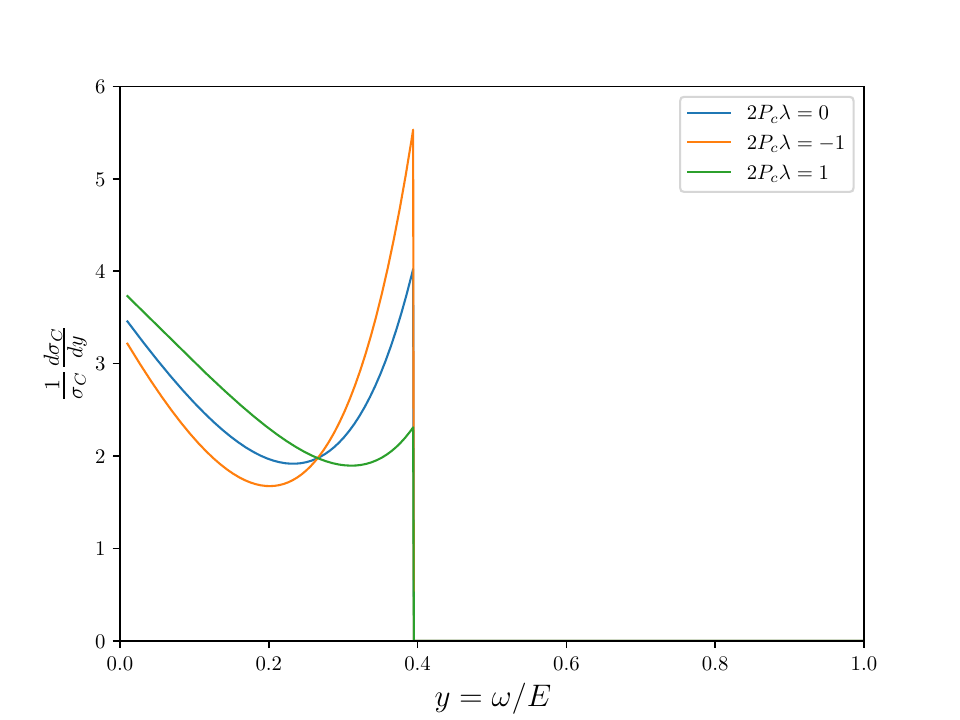} 
	\end{subfigure}
	\begin{subfigure}[b]{0.5\linewidth}
		\centering
		\includegraphics[width=1\linewidth]{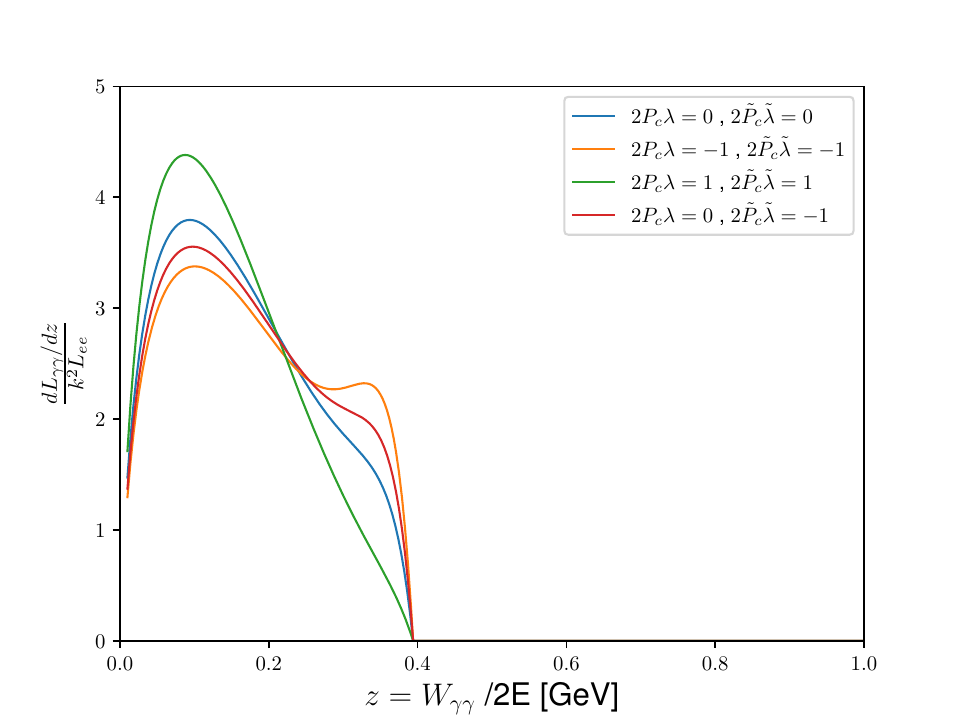} 
	\end{subfigure} 
	\caption{Unpolarized energy spectrum for Compton backscattered photons at different energies (top left) and luminosity spectrum for a gamma-gamma collider from unpolarized electron beams for different $e$-beam energies (top right). Energy spectrum for the photon beam from $17.5$ GeV European XFEL SC electron beam for different helicities (bottom left) and the luminosity spectrum for a gamma-gamma collider for different combinations of helicities of the two electron beams and laser system (bottom right).}
	\label{fig:4plots} 
\end{figure}

The Compton cross section for the backscattering process is given by \cite{GINZBURG2}:
\begin{equation}
	\sigma_c^{np} = \frac{2\pi\alpha^2}{x m_e^2}\left[ \left(\ 1-\frac{4}{x}-\frac{8}{x^2}\right) \ln{(x+1)}+\frac{1}{2}+\frac{8}{x}-\frac{1}{2(x+1)^2}\right],
	\label{eq:compton}
\end{equation}
resulting in the following energy spectrum for the backscattered photons
\begin{equation}
	\begin{split}
		\frac{1}{\sigma_c^{np}}\frac{\text{d}\sigma_c}{\text{d}y} &\equiv f_{np}(x,y) \\ &= \frac{2\pi\alpha^2}{\sigma_c^{np} x m_e^2}\left[ 1-y+\frac{1}{1-y}-\frac{4y}{x(1-y)}+\frac{4y^2}{x^2(1-y)^2}\right].
	\end{split}
	\label{eq:energySpec}
\end{equation}
The energy spectrum, shown in the top left plot of Fig. \ref{fig:4plots}, shows that the broad distribution reaches a maximum at $y=\omega_{max} / E$, but also that a significant amount of photons have energies $\omega\ll\omega_{max}$.
For future colliders the electron beams will have a polarization of up to $80\%$, having drastic impact on the interaction. This also plays a role in Compton backscattering, giving rise to a correction term $\propto 2P_c\lambda$ with $P_c$ and $\lambda$ the mean photon and electron helicity respectively \cite{Ginzburg3},
\begin{equation}
	\sigma_1=\frac{2\pi\alpha^2}{xm_e^2}\left[ \left( 1+\frac{2}{x} \right)\ln{x+1}-\frac{5}{2}+\frac{1}{x+1}-\frac{1}{2(x+1)^2} \right],
	\label{eq:comptonPolcorrection}
\end{equation}
changing the Compton cross section,
\begin{equation}
	\sigma_c = \sigma_c^{np}+2\lambda P_c\sigma_1,
	\label{eq:comptonPol}
\end{equation}
and the scattered photon energy spectrum,
\begin{equation}
	\begin{split}
		\frac{1}{\sigma_c} \frac{d\sigma_c}{dy} &\equiv f(x,y) \\&= \frac{2\pi\alpha^2}{\sigma_c x m_e^2} \left[ \frac{1}{1-y} + 1-y-4r(1-r)-2\lambda P_c rx(2r-1)(2-y) \right],
	\end{split}
	\label{eq:energySpecPol}
\end{equation}
where
\begin{equation}
	r = \frac{y}{x(1-y)}\leq 1.
\end{equation}
The energy spectrum for different values of $2P_c\lambda$ can be seen in the bottom left plot of Fig. \ref{fig:4plots}, with the significant result that for $2P_c\lambda=-1$ photons with energies close to $\omega_{max}$ are favoured while for $2P_c\lambda=1$ they are suppressed. When having two beams converted to real photons via Compton backscattering the luminosity distribution depending on these energy spectra is given by \cite{GINZBURG2}:
\begin{equation}
	\begin{split}
		\frac{1}{k^2L_{ee}}\frac{\text{d}L_{\gamma\gamma}}{\text{d}z} = 2z&\int_{z^2/y_{max}}^{y_{max}}\frac{\text{d}y}{y}f(x,y)f\left(x,\frac{z^2}{y}\right)\\ &I_0\left( \rho^2\sqrt{\left( \frac{y_{max}}{y} -1 \right)\left( \frac{y_{max}y}{z^2}-1 \right)} \right)\\ &\text{exp}\left[ -\frac{\rho^2}{2}\left( \frac{y_{max}}{y}+\frac{y_{max}y}{z^2}-2 \right) \right],
	\end{split}
	\label{eq:lumiGamma}
\end{equation}
where $k$ is the conversion factor and $L_{ee}$ the $ee$-geometric luminosity. The luminosity distribution for the unpolarized and polarized beams can be seen in the right plot of Fig. \ref{fig:4plots} again, showing that the high energy photons are favoured for both beams with $2P_c\lambda=-1$ and a higher low energy contribution for $2P_c\lambda=1$, allowing one to choose whichever is better for a search by just changing $P_c$.

\section{Light-by-Light scattering}
\label{sec:lbyl}
\begin{figure}[h]
	\centering
	\begin{subfigure}{0.3\textwidth}
		\centering
		\includegraphics[width=\linewidth]{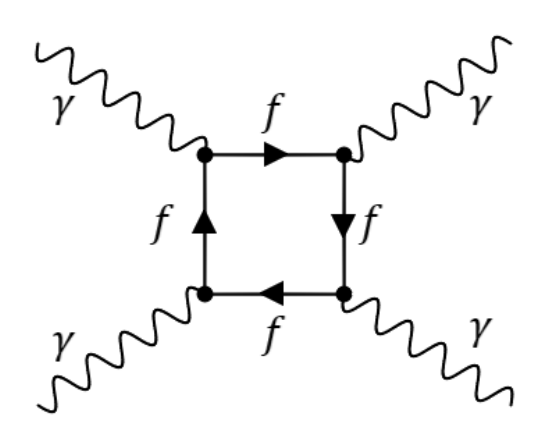}
	\end{subfigure}
	\begin{subfigure}{0.3\textwidth}
		\centering
		\includegraphics[width=\linewidth]{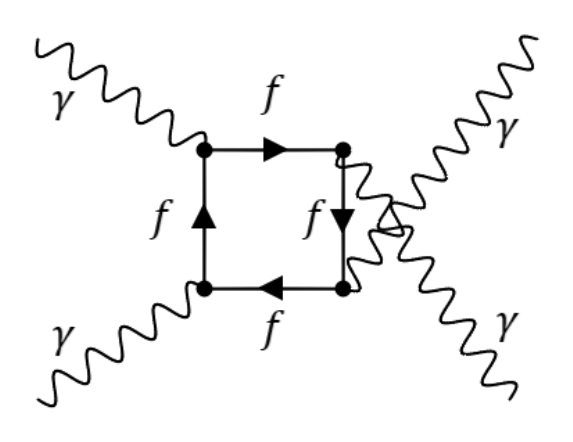}
	\end{subfigure}
	\begin{subfigure}{0.3\textwidth}
		\centering
		\includegraphics[width=\linewidth]{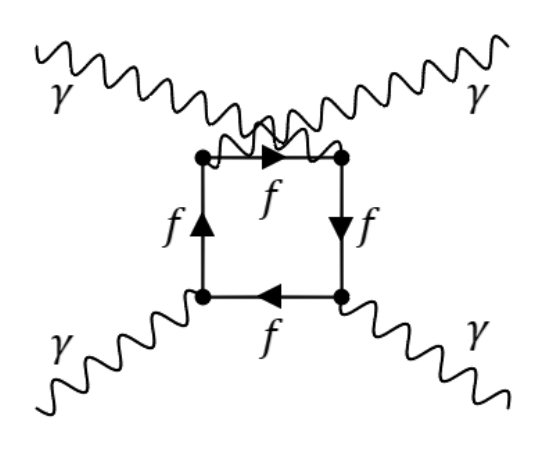}
	\end{subfigure}    
	\caption{The feynman diagrams of light-by-light scattering including fermions.}
	\label{fig:FermionLbyL}
\end{figure}

\begin{figure}[htbp]
	\centering
	\includegraphics[width=0.5\linewidth]{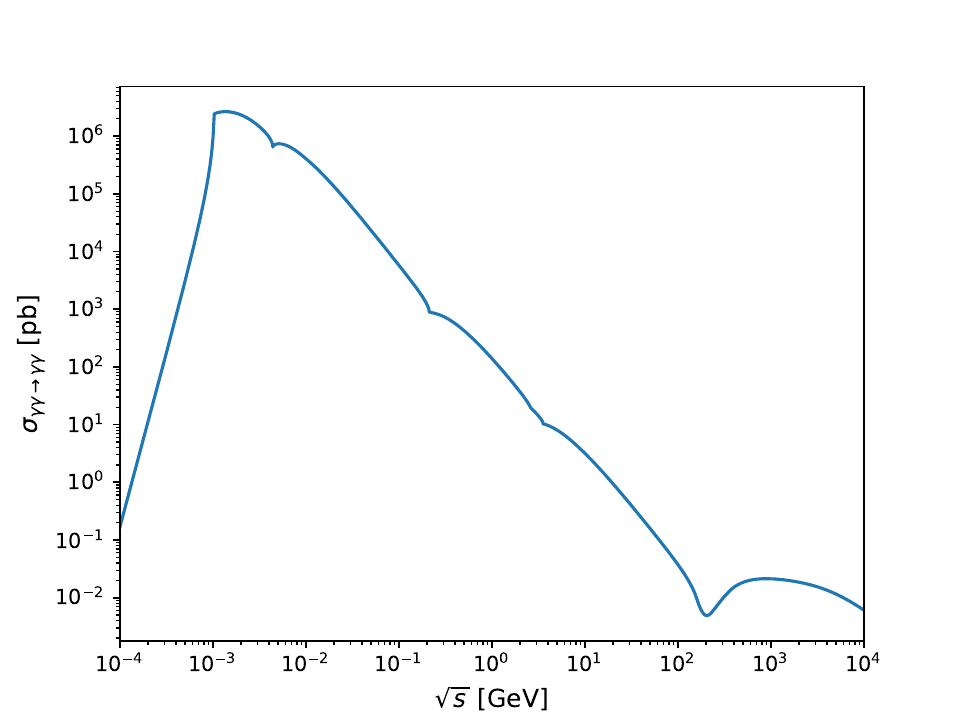}
	\caption{Cross section of light-by-light scattering}
	\label{fig:lbyl}
\end{figure}

The Feynman diagrams contributing to the process $\gamma\gamma\rightarrow\gamma\gamma$ in the SM are shown in Fig. \ref{fig:FermionLbyL}. Here only the fermionic contributions are shown, with the omitted diagrams differing only in the orientation of the fermion lines. 
The diagrams including $W$-bosons are not shown, since they do not contribute at lower energies, but they can be found in Figs. (2,3) in \cite{Boehm} and they will be included in this work for future colliders as well as completeness. We used the \textsc{Mathematica} package $FeynArts$ \cite{FeynArts} to generate and evaluate these diagrams.
The differential cross section in the center of mass frame for this process takes the form 
\begin{equation}
	\frac{\text{d}\sigma}{\text{d}\Omega} = \frac{1}{64\pi^2}\frac{1}{s}|M_{fi}|^2.
	\label{eq:2to2CrossSection}
\end{equation}
In order to evaluate the squared amplitude for light-by-light scattering, it is common to use the helicity amplitudes approach. The resulting number of amplitudes includes all combinations of photon polarizations, therefore a total of $16$ amplitudes would have to be calculated. Making use of $P$-invariance, $T$-invariance and crossing symmetry of this process, we can reduce the number down to three independent amplitudes \cite{Boehm}
\begin{equation}
	M_{++++}(s,t,u), \quad M_{++--}(s,t,u), \quad M_{+++-}(s,t,u).
\end{equation}
Taking the expressions from $FeynArts$ and making use of the in $FormCalc$ \cite{FormCalc} included function $VecSet$ the three amplitudes were evaluated. We found good agreement to expressions given in literature, see \cite{Boehm} eqs. (25-35).
With this we can replace $|M_{fi}|^2$ in the differential cross section eq. (\ref{eq:2to2CrossSection}) by
\begin{equation}
	|M_{fi}|^2 \rightarrow \frac{1}{2}(|M_{++++}|^2+|M_{++--}|^2+|M_{+-+-}|^2+|M_{+--+}|^2+4|M_{+++-}|^2),
	\label{eq:SquaredAmplitudes}
\end{equation}
where 
\begin{equation}
	\begin{split}
		M_{+-+-}(s,t,u) &= M_{++++}(u,t,s),\\
		M_{+--+}(s,t,u) &= M_{++++}(t,s,u),
		\label{eq:crossingsym}
	\end{split}
\end{equation}
from crossing symmetry, leading to the differential cross section:
\begin{equation}
	\frac{d\sigma}{d\Omega} = \frac{1}{128\pi^2 s} (|M_{++++}|^2+|M_{+-+-}|^2+|M_{+--+}|^2+|M_{++--}|^2+4|M_{+++-}|^2).
	\label{eq:differentialHeli}
\end{equation}
The total unpolarized cross section for light-by-light scattering $\sigma_{\gamma\gamma\rightarrow\gamma\gamma}$ is shown in Fig. \ref{fig:lbyl}, the bumps in the graph representing the contributions of different particles. At lower energies $\sqrt{s} < 2 m_W$ the overall shape is dominated by the electron contribution, whereas the other fermions only shifting the cross section slightly, corresponding to the bumps. At higher energy $\sqrt{s} > 2m_W$ the cross section mainly depends on the $W$-boson contribution while the fermionic contribution is negligible. 

In the energy range $W_{\gamma\gamma} < 12 \text{ GeV}$ reachable by the discussed gamma-gamma collider based on European XFEL, the cross section is still of order pb, therefore it would be very probable to achieve good results for this process. For future colliders, the strong $W$-boson contribution maintains the cross section of order fb, giving again great observation possibilities.

As shown in the previous chapter, the luminosity distribution for a gamma-gamma collider is not a delta peak but instead also has large contributions from lower energies. We used eqs. (\ref{eq:energySpec}), (\ref{eq:energySpecPol}) to calculate the photon energy spectra. Using 
\begin{equation}
	\text{d}\sigma = 2\int_{0}^{z_{max}}\text{d}z \text{ }z\int_{z^2/y_{max}}^{y_{max}}\text{d}y\text{ }\frac{1}{y}f(x,y)f\left(x,\frac{z^2}{y}\right) \text{d}\sigma^{\gamma\gamma\rightarrow\gamma\gamma},
	\label{eq:GammaGamma}
\end{equation}
we can calculate the cross section for the corresponding process at a gamma-gamma collider. Using the parameters proposed for the gamma-gamma collider with energies < $12$ GeV based on European XFEL \cite{Telnov_2020}, the calculated cross sections are shown in Tab. \ref{tab:lbyl} for light-by-light scattering.

\begin{table}[htbp]
	\centering
	\begin{tabular}{c|c|c|c}
		$2 \lambda P_c$ & $2 \tilde{\lambda} \tilde{P}_c$ & $\sigma_{\gamma\gamma\rightarrow\gamma\gamma}$ [pb] & $\sigma_{\gamma\gamma\rightarrow\gamma\gamma}^{W_{\gamma\gamma}>6\text{ GeV}}$ [pb] \\
		\hline
		0 & 0 & 74.72 & 1.528\\
		-1 & -1 & 65.02 & 1.581\\
		1 & 1 & 86.55 & 1.429\\
		0 & -1 & 69.7 & 1.563
	\end{tabular}
	\caption{Cross sections for light-by-light scattering at a gamma-gamma collider at the European XFEL for different values of the average electron $2\lambda$ and laser photon $P_c$ helicities, where $\sim$ denotes the second beam. The right cross section has a cut applied to the center of mass energy of $W_{\gamma\gamma}>6$ GeV. The polarization of the colliding photon beams was hereby not taken into consideration.}
	\label{tab:lbyl}
\end{table}
The difference in these total cross sections again shows the luminosity spectrum dependency as shown in Fig. \ref{fig:4plots}.. Without the cut, the highest cross section would be measured for both beams having $2P_c\lambda = 1$, as the curve has high contributions from the low energy region, scaling better with the shape of Fig. \ref{fig:lbyl}. But when applying the cut on $W_{\gamma\gamma} > 6$ GeV ($z\approx 0.17$), the highest cross section would be measured for both beams with $2P_c\lambda = -1$, again corresponding to the higher contribution in Fig. \ref{fig:4plots} at this energy range.

\section{ALPs}
\label{sec:alps}
\begin{figure}[h]
	\centering
	\begin{subfigure}{0.3\textwidth}
		\centering
		\includegraphics[width=\linewidth]{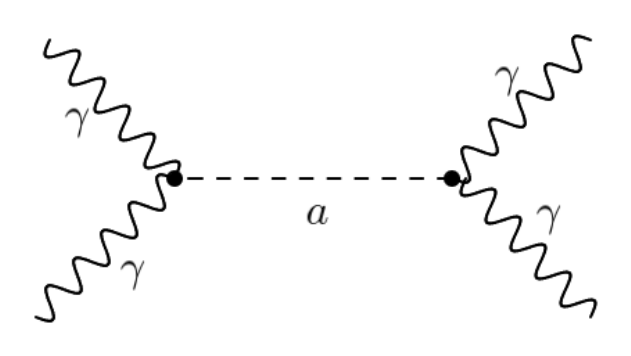}
	\end{subfigure}
	\begin{subfigure}{0.3\textwidth}
		\centering
		\includegraphics[width=\linewidth]{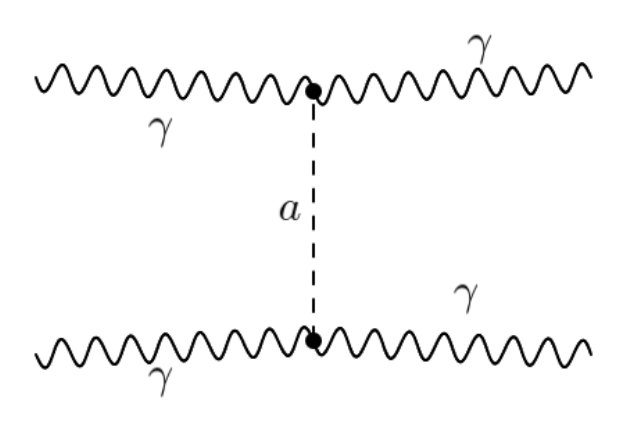}
	\end{subfigure}
	\begin{subfigure}{0.3\textwidth}
		\centering
		\includegraphics[width=\linewidth]{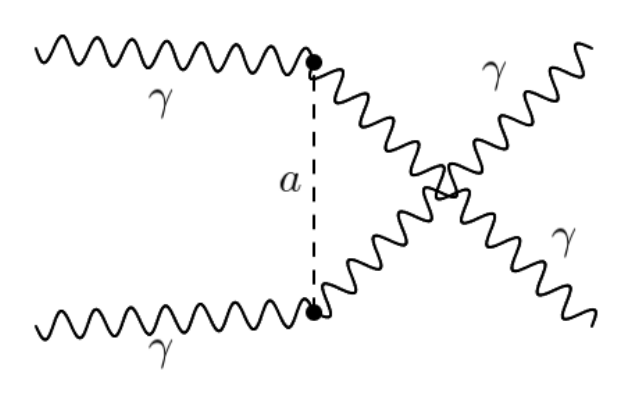}
	\end{subfigure}    
	\caption{Feynman diagrams for ALP-production in the $\gamma\gamma\rightarrow\gamma\gamma$ channel.}
	\label{fig:ALPproduction}
\end{figure}
\begin{figure}[h]
	\centering
	\begin{subfigure}{0.4\textwidth}
		\centering
		\includegraphics[width=\linewidth]{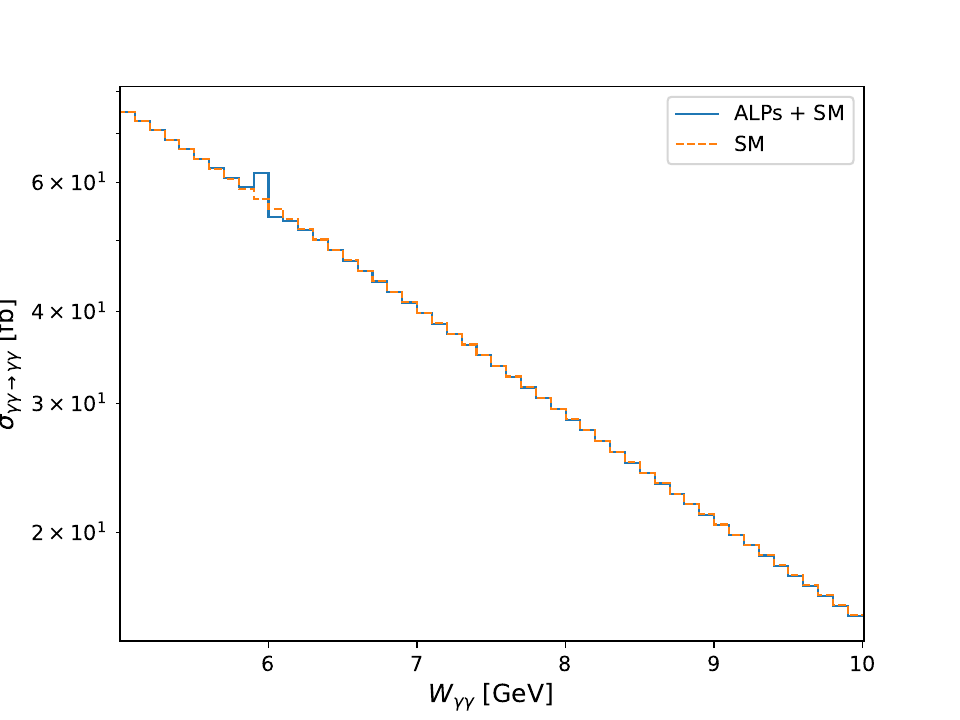}
	\end{subfigure}
	\begin{subfigure}{0.4\textwidth}
		\centering
		\includegraphics[width=\linewidth]{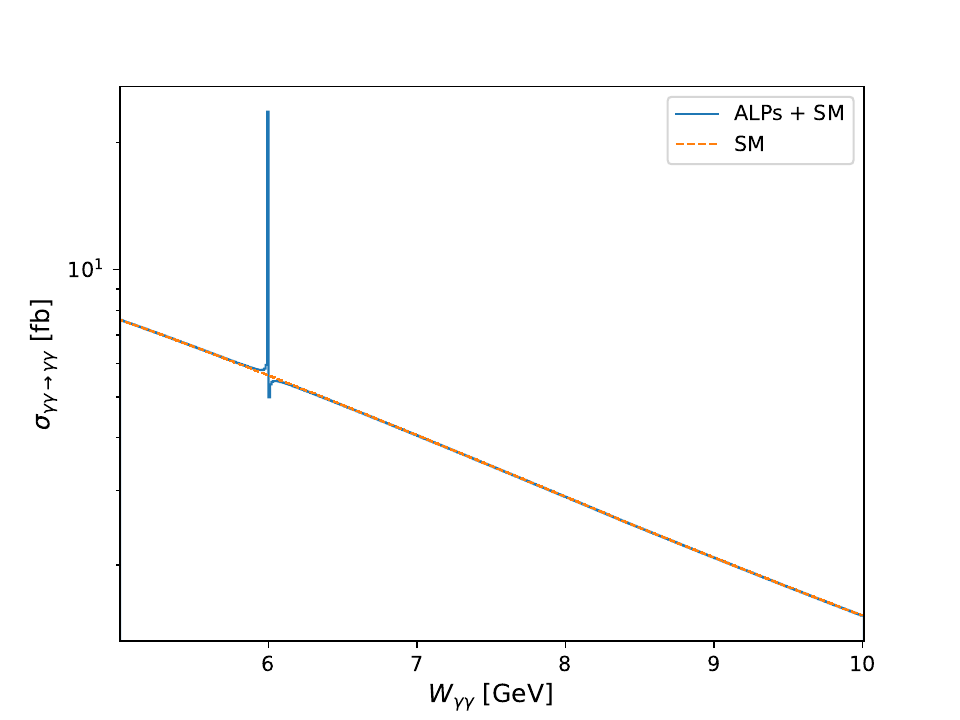}
	\end{subfigure}    
	\caption{Cross section for the process $\gamma\gamma\rightarrow a\rightarrow\gamma\gamma$ compared to the pure SM process over the photon invariant mass $W_{\gamma\gamma}$. In the left plot the bin size has been chosen to be $100$ MeV and in the right plot $10$ MeV.}
	\label{fig:alpsCross}
\end{figure}

The ALP coupling only to photons is described by
\begin{equation}
	\mathcal{L}_{ALPs}=\frac{1}{2}\partial^\mu a \partial_\mu a-\frac{1}{2}m_a^2 a^2-\frac{a}{f_a}F_{\mu\nu}\Tilde{F}^{\mu\nu},
\end{equation}
where $a$ is the axion field, $m_a$ is the mass of the axion, $F_{\mu\nu}$ is the electromagnetic field strength tensor, with $\Tilde{F}^{\mu\nu}=\frac{1}{2}\epsilon^{\mu\nu\rho\sigma}F_{\rho\sigma}$ and $f_{a}$ the ALP-photon coupling. Following from this, the differential cross section of the process $\gamma\gamma\rightarrow a\rightarrow\gamma\gamma$, the respective Feynman diagrams are shown in Fig. \ref{fig:ALPproduction}, can be evaluated again by eq. (\ref{eq:SquaredAmplitudes}) and eq. (\ref{eq:differentialHeli}), with $M_{\lambda_1\lambda_2\lambda_3\lambda_4}$ given as the sum of the ALP and SM terms,
\begin{equation}
	M=M_{ALP}+M_{SM}.
\end{equation}
The ALP amplitudes are given as 
\begin{align}
	M_{++++}^{ALP}(s,t,u) &= \frac{4}{f_a^2} \frac{s^2}{s-m_a^2+im_a \Gamma_a},\label{eq:ALP++++}\\
	M_{++--}^{ALP}(s,t,u) &= \frac{4}{f_a^2} \left(\frac{s^2}{s-m_a^2+i m_a \Gamma_a} + \frac{t^2}{t-m_a^2} + \frac{u}{u-m_a^2} \right),\label{eq:ALP++--}\\
	M_{+-+-}^{ALP}(s,t,u) &= M_{++++}^{ALP}(u,t,s)= \frac{4}{f_a^2} \frac{u^2}{u-m_a^2},\label{eq:ALP+-+-}\\
	M_{+--+}^{ALP}(s,t,u) &= M_{++++}^{ALP} (t,s,u) = \frac{4}{f_a^2} \frac{t^2}{t-m_a^2},\label{eq:ALP+--+}\\
	M_{+++-}^{ALP}(s,t,u) &=0,\label{eq:ALP+++-}             
\end{align}
with
\begin{equation}
	\Gamma_a=\frac{1}{\text{Br}(a\rightarrow\gamma\gamma)}\frac{m_a^3}{4\pi f_a^2}
	\label{eq:width}.
\end{equation}
The ALP width $\Gamma_a$ has been omitted in the eqs. (\ref{eq:ALP+-+-}), (\ref{eq:ALP+--+}) and in the last two terms of eq. (\ref{eq:ALP++--}) due to the negligible impact away from $s\sim m_a^2$. 
The cross section for ALP production at the discussed gamma-gamma collider can be seen in Fig. \ref{fig:alpsCross}. Here an ALP mass of $m_a=6$ GeV and a coupling constant of $f_a =10$ TeV  was chosen, to look at a specific scenario. It can be seen that over the dependence on $W_{\gamma\gamma}$ a peak-dip structure is measurable, but due to the very small width (\ref{eq:width}) when integrating over the whole energy range, the result would be indistinguishable from the SM light-by-light scattering. The bins have been chosen to be $100$ MeV and $10$ MeV wide, this would be very challenging for current detector technology, but with a high enough resolution it would be possible to see the structure and differentiate between the pure SM light-by-light scattering and an extension with an ALP with a coupling constant of this order. 

\section{Conclusion and outlook}
\label{sec:outlook}
The unique advantage of extending future $e^+e^-$ linear colliders with a gamma-gamma collider via Compton backscattered photons has been discussed, giving access to $\gamma\gamma$ and $\gamma e$ interactions. The concept has been proposed many years ago, but no gamma-gamma collider has been realized until today, therefore it would be advantageous to develop the expertise and the knowledge at current facilities, so that once the future colliders are ready to be build the full potential can be used right away. For this one could use an existing electron beam, like the $17.5$ GeV electron beam of the European XFEL SC to build the first gamma-gamma collider with the laser system optimal for future high energy $e^+e^-$ collider. This possible setu-up has been used to calculate the light-by-light scattering cross section in the pure SM as well as the extension via an ALP. These two processes have been compared, showing that it would be possible to differentiate between them, in case a sufficiently high enough photon energy resolution can be achieved.\\
Further results for future $e^+e^-$ colliders such as ILC and HALHF, are under work using codes like CAIN for simulations of the corresponding luminosity spectra.

\section*{Acknowledgements}

We would like to thank V. I. Telnov and K. Yokoya for many interesting discussions and providing input for the usage of CAIN. This work is being supported by the Deutsche Forschungsgemeinschaft (DFG, German Research Foundation) under MO-2197/2-1.

\bibliography{bibliography.bib}

\end{document}